# Electrically generated exciton polaritons with spin on-demand

Yutao Wang,[1,2,3] Giorgio Adamo,[1,2] Son Tung Ha,[4] Jingyi Tian,[1,2,†] and Cesare Soci[1,2,*]

[1] *Centre for Disruptive Photonic Technologies, TPI, Nanyang Technological University, 21 Nanyang Link, Singapore 637371*

[2] *Division of Physics and Applied Physics, School of Physical and Mathematical Sciences, Nanyang Technological University, 21 Nanyang Link, Singapore 637371*

[3] *Interdisciplinary Graduate School, Energy Research Institute @NTU (ERI@N), Nanyang Technological University, 50 Nanyang Drive, Singapore 637553*

[4] *Institute of Materials Research and Engineering, Agency for Science Technology and Research (A*STAR), 2 Fusionopolis Way, Singapore 138634*

[†]*Current address*: *School of Engineering, Westlake University, Hangzhou, Zhejiang, China 310024*

**Correspondence:* csoci@ntu.edu.sg

**Generation and manipulation of exciton polaritons with controllable spin could deeply impact spintronic applications, quantum simulations, and quantum information processing, but is inherently challenging due to the charge neutrality of the polariton and the device complexity it requires. In this work, we demonstrate electrical generation of spin-polarized exciton polaritons in a monolithic dielectric perovskite metasurface embedded in a light-emitting transistor. A finely tailored interplay of in- and out-of-plane symmetry breaking of the metasurface allows to lift the spin degeneracy through the polaritonic Rashba effect, yielding high spin purity with normalized Stokes parameter of $S_3 \sim 0.8$. Leveraging on spin-momentum locking, the unique *metatransistor* device architecture enables electrical control of spin and directionality of the polaritonic emission. This work advances the development of compact and tunable spintronic devices, and represents an important step toward the realization of electrically pumped inversionless spin-lasers.**

Exciton-polaritons are hybrid quasiparticles of bosonic nature that emerge from the strong coupling between confined electromagnetic modes and electron-hole bound states in condensed matter. As such, the properties they inherit from their photonic and electronic constituents, that is small effective mass, strong nonlinearities and fast relaxation,[1-9] make them an exciting platform for fundamental studies of Bose-Einstein condensation[1,3,10], superfluidity[5,11], and for the realization of classical[12-14] and quantum optoelectronic devices[15]. Imparting an additional spin degree of freedom to the polaritons, either through the electronic or the photonic spin, can open new avenues for spintronic applications: the ability to efficiently create highly pure spin-polariton states, to electrically generate them, and to dynamically manipulate spin selectivity and directionality, without the need of external optical components, would have a vast impact for the realization of functional opto-spintronic devices[16-19], quantum information processors[15] and neuromorphic computers[8]. However, the realization of electrically driven spin-polarized polariton devices has been hindered so far by the inherent charge neutrality of the polaritons and the complex device architectures needed to control the polariton spin.

Due to their bosonic nature, the spin of polaritons can be described by an integer number ($\sigma = \pm 1$). The creation of a spin-polarized polaritonic states requires lifting the spin degeneracy. While this is attainable by using magnetic fields to break time-reversal symmetry,[20-22] magnetism-free methods for producing spin-polarized polaritons are actively researched. One possibility is to use excitonic materials with broken inversion symmetry in their electronic transitions embedded in microcavities.[23-26] A more flexible approach, unbound by the electronic spin properties of the material, is to introduce spin-orbit coupling in photonic microcavities[17-19,27,28]. This approach has led to fascinating implementations of polaritonic topological insulators[28], synthetic gauge fields[27], and polaritonic spin Hall effects[17].

Recently, photonic bound states in the continuum (BIC) - modes that remain localized in spite of existing in a continuous spectrum of radiating waves,[29,30] have been used to realize perfectly confined states with theoretically infinite-Q resonances, facilitating strong light-matter interaction.[10,31-34] Γ-BIC, which are protected by the symmetry mismatch between the photonic mode profiles and the out-of-plane propagating waves, manifest as polarization singularities (V-points) at the centre of the Brillouin zone. Due to the spin-orbit coupling, Γ-BIC can be evolved into off-Γ, purely circularly polarized states (C-points) by breaking the in-plane inversion symmetry of the photonic cavity (a phenomenon also referred to as the optical Rashba effect[35-38], in analogy with the condensed matter Rashba effect[39-41]). Usually, this comes at the expense of lowering the Q-factor of the modes, weakening the coupling strength at C-points.

Here we demonstrate the electrical generation and control of highly spin-polarized exciton polaritons in a perovskite light-emitting metatransistor. The device combines, in a monolithic structure, a dielectric metasurface designed to break the in-plane inversion symmetry and a dielectric stack that breaks the out-of-plane inversion symmetry, resulting in purely circularly polarized resonances with high Q-factors. By supporting strong coupling of the highly confined excitons in the perovskite, this system yields exciton polaritons with spin purity approaching the theoretical limit. Upon charge injection in the transistor channel, we observe electroluminescence characterized by a marked *polaritonic* Rashba effect (the spin-split of polariton bands in opposite halves of the momentum space). Moreover, we show that individual polaritonic spin states can be selected electrically by controlling charge (exciton) injection, and thus imparting specific directionality and helicity to the circularly polarized electroluminescence (EL).

**Device concept**

The electrically driven light emitting metasurface adopts a typical top-contact, bottom-gate light emitting transistor (LET) configuration, so that the excitonic emission channel and the metasurface pattern can be co-located between the source and drain electrodes[42]. The metasurface is directly fabricated into the $MAPbI_3$ film with high uniformity, by focused ion beam (FIB) lithography. The LET configuration enables operation of the device in the ambipolar transport regime under certain electrical biasing conditions, where electrons and holes are simultaneously injected from the source and the drain electrodes into the transistor channel at the semiconductor-insulator interface, and form excitons that radiatively recombine to emit light[43,44]. Moreover, tuning of the relative bias between source and drain electrodes enables control of the lateral position of the recombination zone within the transistor channel[45], a unique feature which here we exploit to implement electrical control of the directional energy flow of exciton-polaritons and their spin in symmetry-broken BIC metasurfaces operating in the strong coupling regime (Fig. 1).

In practice, our light-emitting *metadevices* can operate in two modes: i) generate EL emission from exciton polaritons of opposite spin, as prescribed by the polaritonic Rashba effect (Figs. 1a and 1b); ii) emit unidirectional light stemming from exciton polaritons with spin-down (Figs. 1c and 1d) or spin-up (Figs. 1e and 1f) into the left or the right halves of the outcoupling hemisphere. Fig. 1a illustrates the optical dispersion of the passive photonic mode in the broken symmetry metasurface that lifts the spin degeneracy. The photonic mode encompasses spin-up states for $k_x > 0$ and spin-down states for $k_x < 0$ in the opposite halves of the momentum space (coloured in red and blue, respectively). Upon strong coupling, the anticrossing of the photonic mode and the perovskite exciton leads to the emergence of exciton-polariton bands (UPB and LPB), inheriting the spin properties of the photonic band. When the transistor is operated under positive drain bias ($V_d > 0$), excitons are generated uniformly within the metasurface and exciton polaritons with opposite spins that decay radiatively outcouple from the metasurface in opposite directions (Fig. 1b). When photons and polaritons of positive (negative) group velocity $u_x = (1/\hbar)dE/dk_x$ propagate in the metasurface, the spin-down (spin-up) branch is selectively populated, as shown in Fig. 1c (Fig. 1e). Combined with spin-momentum locking, the change of group velocity within the metasurface induced by negative bias applied to either the drain ($V_d < 0$) or source ($V_s < 0$) electrodes, enables electrical control of helicity and directionality of the light emitted by spin-polarized polaritons (Figs. 1d and 1f).

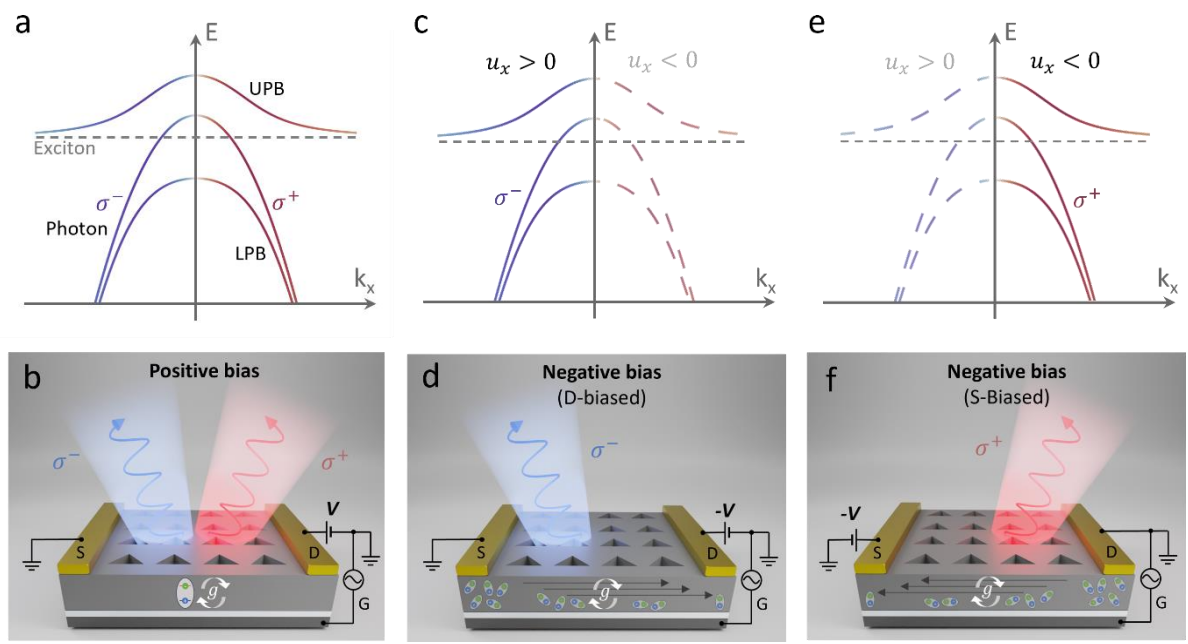

**Figure 1. Tunable spin-polarized exciton-polaritons in the electrically driven perovskite metatransistor.** The light-emitting metatransistor operates in two distinct modalities: i) generates circularly polarized EL from exciton polaritons of opposite spin, as prescribed by the polaritonic Rashba effect; ii) unidirectionally emits circularly polarized light from spin-down and spin-up exciton polaritons with either positive or negative group velocities. a) Characteristic band diagram of a metasurface with broken in-plane inversion symmetry, for the case of passive photonic modes and exciton-polariton bands, where the spin degeneracy is lifted due to the optical and polaritonic Rashba effect, respectively. The red and blue colours denote spin-up and spin-down states, respectively; b) Schematic of Rashba polariton emission from the electrically driven metasurface under positive drain

bias, operating in modality i). c, e) Dispersion of the optical bands with lifted spin degeneracy featuring positive ($u_x > 0$) and negative group velocities ($u_x < 0$), both for photon and polariton modes; bands with unmatched group velocities are not populated (dashed lines). d, f) Schematics of spin polarized emission from the device operating in modality ii), with negative drain bias (D-biased condition) or negative source bias (S-biased condition). Spin-momentum locking and asymmetric exciton generation enable electrical control of directionality and helicity of the spin polarized polaritonic emissions.

**Strong coupling in high-Q perovskite metasurface with broken inversion symmetry**

Methylammonium lead iodide (MAPbI$_3$) perovskite was selected for this proof-of-principle demonstration of an electrically-driven spin-polarized polariton light-emitting device because it combines excellent excitonic and charge transport properties[46-49]. Its high refractive index also allows for monolithic fabrication of metasurfaces[50-52] in light-emitting transistor devices. Absorption measurements of a pristine MAPbI$_3$ film at 78 K yield exciton energy of $E_{exc}$ = 1.693 eV and damping rate of $\gamma_{exc}$ = 68 meV using the Elliot model fitting (Supplementary Note 1 and Fig. S1). The exciton binding energy extracted from the model ($E_b$ ~30 meV) is larger than the thermal energy even at room temperature, fulfilling the prerequisite for the formation of stable exciton-polaritons[46]. The low temperature electrical characteristics of the MAPbI$_3$ perovskite reveal slightly unbalanced ambipolar charge carrier injection, with higher mobility observed for electrons (Supplementary Note 2 and Fig. S2).

Our monolithic metasurface design uses an isosceles triangular hole unit cell, with broken in-plane inversion symmetry (Fig. 2a and 2b)[36,37,53]. The optical response of the perovskite metasurface was numerically evaluated by modelling the refractive index of MAPbI$_3$ as a Tauc-Lorentz dielectric, both in the absence and presence of excitons (Supplementary Note 3 and Fig. S3). The band structure of the passive perovskite metasurface in the absence of excitons, calculated by eigenmode simulations, displays a pair of pure $\sigma^+$ and $\sigma^-$ states ($|S_3| = 1$) adjacent to the Γ-point on the Ph$_2$ band, as seen in Fig. 2c. These circularly polarized states, known as C-points, emerge from the splitting of the polarization singularity (V-point) of the symmetry-protected BIC upon breaking the in-plane inversion symmetry of the metasurface. Conventionally, in a triangular hole metasurface, where only the in-plane IS is broken, the Q-factor for the Ph$_2$ band peaks at the Γ-point. Here the mode is linearly polarized in the far field ($|S_3| = 0$) and the Q-factor strongly decreases toward the C-points, as shown in the upper panel of Fig. 2d. Recently, it was shown that high Q-factor points can be created anywhere off Γ exploiting the hybridization of photonic resonances when the out-of-plane symmetry of the system is broken[54]; alternatively, one of the C-points could be shifted to the high-Q Γ-point through a slant along the vertical axis of the metasurface unit cell[55,56]. Here we make use of the transistor stack, which breaks the out-of-plane symmetry by design (Fig. 2a), and optimize the thickness of the SiO$_2$ dielectric layer (500 nm) between the Si substrate and the perovskite film to produce two high Q-factor modes at the C-points (at $k_x = \pm 0.125$ µm$^{-1}$) on the Ph$_2$ band (lower panel of Fig. 2d). The creation of these high-Q points stems from the interplay between Ph$_1$ and Ph$_2$ bands, which would be nearly degenerate in the absence of vertical symmetry breaking. A detailed investigation of the symmetry-dependent Q factor for the modes in the Ph$_2$ band can be found in Supplementary Note 4, Fig. S4). It is worth noting that our design strategy allows enhancing the Q-factor of both circularly polarized photonic modes without requiring additional layers or complex fabrication steps, offering a simple route to achieve spin-polarized polaritonic emission from electrically driven devices.

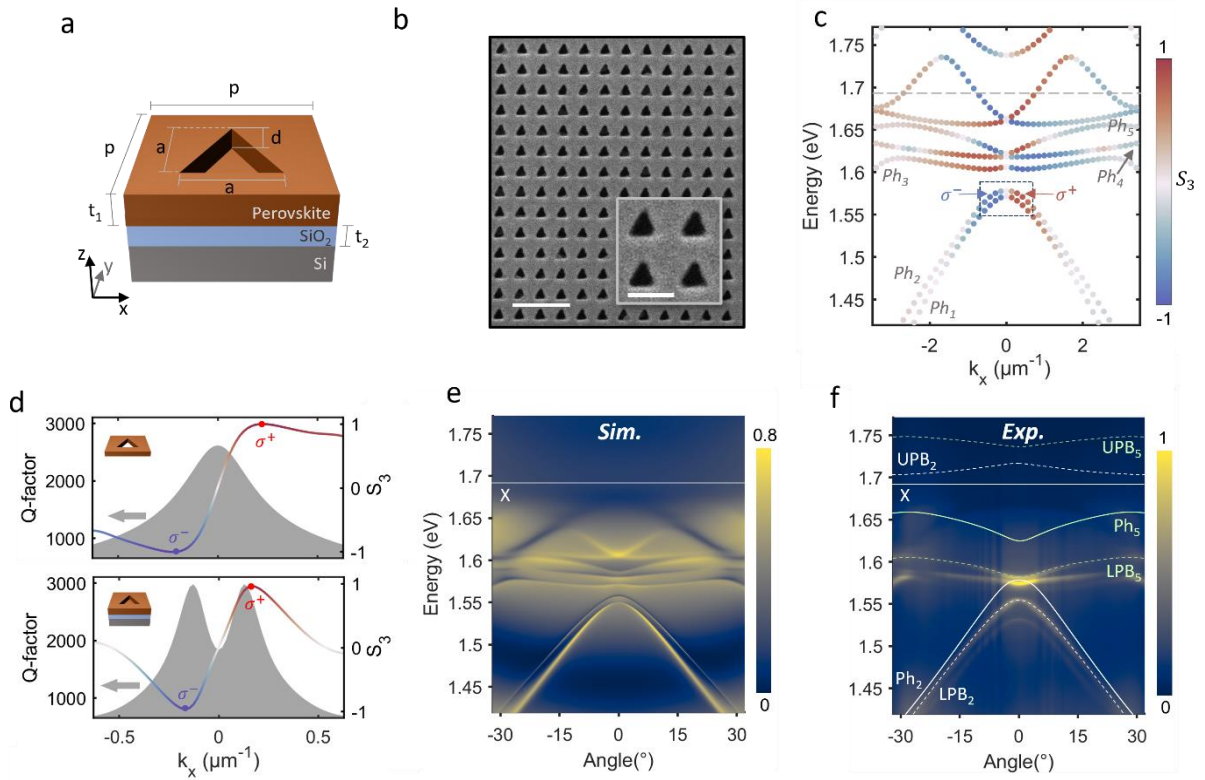

**Figure 2. Engineering strong chiral light-matter interaction in perovskite metasurfaces with broken inversion symmetry**. a) Schematic of the quasi-BIC unit cell on a SiO$_2$-Si substrate. The geometric parameters are set as follows: period p = 355 nm, base and height of the triangular hole a = 0.6×p, thickness of the perovskite film $t_1$ = 350 nm, depth of the triangle hole d = 250 nm, thickness of the SiO$_2$ layer $t_2$ = 500 nm. b) Scanning electron microscope image of the fabricated perovskite metasurface. The scale bar is 1 µm. The inset shows an enlarged view of four unit cells. The scale bar is 300 nm. c) Numerically simulated passive photonic bands of the perovskite metasurface with broken in-plane and out-of-plane inversion symmetry (IS). The colormap indicates the value of the normalized Stokes parameter $S_3$. The grey dashed line denotes the energy of the excitonic mode of the MAPbI$_3$ perovskite. d) Simulated Q-factor (shaded areas) and $S_3$ (colored lines) of the photonic Ph$_2$ band for a passive perovskite metasurface with only the in-plane IS breaking (upper panel) and both in-plane and

out-of-plane IS breaking (lower panel), showing the emergence of two high-Q peaks in correspondence of the C-points (blue and red dots). The insets show the schematic of the two designs. e) Simulated angle-resolved reflection spectra of the active perovskite metasurface, when accounting for the MAPbI$_3$ exciton. The white solid line denotes the excitonic mode of perovskite (X). f) Experimental angle-resolved reflection spectra of the perovskite metasurface. The white and green solid lines illustrate the dispersion of two passive photonic modes (Ph$_2$ and Ph$_5$ bands). The white and green dashed lines denote the dispersion of the corresponding polariton bands for Ph$_2$ (LPB$_2$, UPB$_2$) and Ph$_5$ (LPB$_5$, UPB$_5$), respectively.

Excitonic states with energy $E_{exc}$ within the monolithic metasurface coherently couple to spin-momentum locked photonic modes with good spectral overlap (Fig 1b). The energy of the resulting upper and lower polariton states can be calculated according to the coupled oscillator model[3]:

$$E_{UP,LP} = \tfrac{1}{2}(E_{cav} + i\gamma_{cav} + E_{exc} + i\gamma_{exc}) \pm$$
$$\sqrt{4g^2 + [E_{exc} + i\gamma_{exc} - E_{cav} + i\gamma_{exc} - i\gamma_{cav}]^2} \qquad (1)$$

where $E_{cav}$ and $\gamma_{cav}$ are the energy and the damping rate of the metasurface photonic modes, while $E_{exc}$ and $\gamma_{exc}$ the energy and the damping rate of the excitonic mode of the perovskite. $g$ is the coupling strength between the two modes. The corresponding Rabi splitting energy is $\Omega_R = 2\sqrt{g^2 - \frac{(\gamma_{cav}-\gamma_{exc})^2}{4}}$, which corresponds to the smallest energy difference between the upper and lower polariton bands. The simulated angle-resolved reflection spectra of the active perovskite metasurface (in the presence of excitons), plotted in Fig. 2e, show excellent agreement with the experimental angle-resolved reflection spectra of Fig. 2f. The white and green solid lines in Fig. 2f illustrate the dispersion of the photonic modes Ph$_2$ and Ph$_5$ in the passive metasurface (see Fig. 2c), whereas the white and green dashed lines show the dispersion of the polaritonic bands formed by the MAPbI$_3$ exciton coupled to the photonic modes Ph$_2$ (LPB$_2$, UPB$_2$) and Ph$_5$ (LPB$_5$, UPB$_5$), respectively. The best fit to Eq. 1 yields Rabi splitting energies of the polaritonic bands associated to Ph$_2$ and Ph$_5$ of $g_{Ph2}$ =116 meV and $g_{Ph5}$ =138 meV. The difference between the two can be attributed to the detuning of the photonic modes ($\Delta_{Ph2}$= -114 meV, $\Delta_{Ph5}$= -71 meV). The Rabi splitting energy of polaritonic bands fulfil the criteria $\Omega_R > \gamma_{exc} + \gamma_{cav}$ i.e., $g > \sqrt{(\gamma_{exc}^2 + \gamma_{cav}^2)/2}$. For our two sets of polaritonic bands, the linewidths of the photonic modes obtained from the simulated spectra are $\gamma_{cav_{Ph2}} = 0.5$ meV and $\gamma_{cav_{Ph5}} = 0.6$ meV, thus confirming that the system is in a strong coupling regime. The absence of the higher energy upper polariton bands (above 1.693 eV) in both the simulated and experimental spectra is due to the strong absorption of the perovskite above the band edge. The angle-resolved photoluminescence measurement upon non-resonant optical pumping, presented in Fig. S7, demonstrates the substantial occupation of spin-polarized polaritons in the LPBs, as predicted by the coupled oscillator model.

**Electrically driven polaritonic Rashba effect**

We exploit the light-emitting transistor functionality to control charge injection into the perovskite metasurface and induce strong coupling of the excitons with designer photonic modes. To improve brightness and uniformity of the electroluminescence emitted from the transistor channel, the gate bias is AC-modulated while the balance of electron and hole injection is controlled by DC source-drain bias[57]. Under positive bias conditions ($V_s$ = 0 V, $V_d$ = 90 V and $|V_g|$ = 90 V), we observe a uniform distribution of radiative excitons across the metasurface, as seen in Supplementary Fig. S8a. The measured angle-resolved electroluminescence (EL) spectra shown in Fig. 3a reveal intense emission from the LPBs, in good agreement with calculations (dashed line). The monolithic device architecture enables efficient light matter interaction, so that the signal from the uncoupled excitons is barely visible in the spectra. As spin-polarized polaritons decay as circularly polarized photons and outcouple from the metasurface inheriting the same energy and momentum, the spin-dependent polaritonic band structure can be determined from circularly polarized EL measurements. The spin properties of the polaritons are characterized by the normalized Stokes parameter, $S_3 = \frac{I_{\sigma^+} - I_{\sigma^-}}{|I_{\sigma^+} + I_{\sigma^-}|}$, which can be mapped in the k-space. Angle-resolved EL spectra of opposite polarization handedness $\sigma^+$ (Fig. 3b) and $\sigma^-$ (Fig. 3c) reveal the polaritonic Rashba effect, i.e. the splitting of the electrically-driven spin-polarized exciton polaritons, with $\sigma^+$ and $\sigma^-$ LPBs appearing as mirror images in k-space. The corresponding $S_3$ map, shown in Fig. 3d, indicates a very high degree of circular polarization, reaching a maximum of approximately 0.8 in the LPB$_2$ near ± 2°, a value remarkably close to the calculated $|S_3|$ = 1, when accounting for imperfections induced by the metasurface fabrication. As discussed in Fig. 2d, we observe enhanced Q-factors that peak near the C-points in the EL spectra, revealed by the reduced linewidth (Fig. S6) and the ~11-fold maximum enhancement of the EL from the LPB$_2$ at the same angle. Figs. 3e-g illustrate the total, right circularly polarized, and left circularly polarized far-field profiles of the polaritonic emission from LPB$_2$ in the vicinity of the Γ-point, collected using a 10-nm bandwidth (~20 meV) bandpass filter centred around 1.55 eV. The corresponding $S_3$, shown in Fig. 3h, reveals the high purity of the spin-polarized polaritons, with an integrates $|S_3|$ value reaching ~0.7, consistent with the anisotropy map in Fig. 3d. Additional field patterns of the Rashba polaritons obtained at different EL energies are illustrated in Fig. S8.

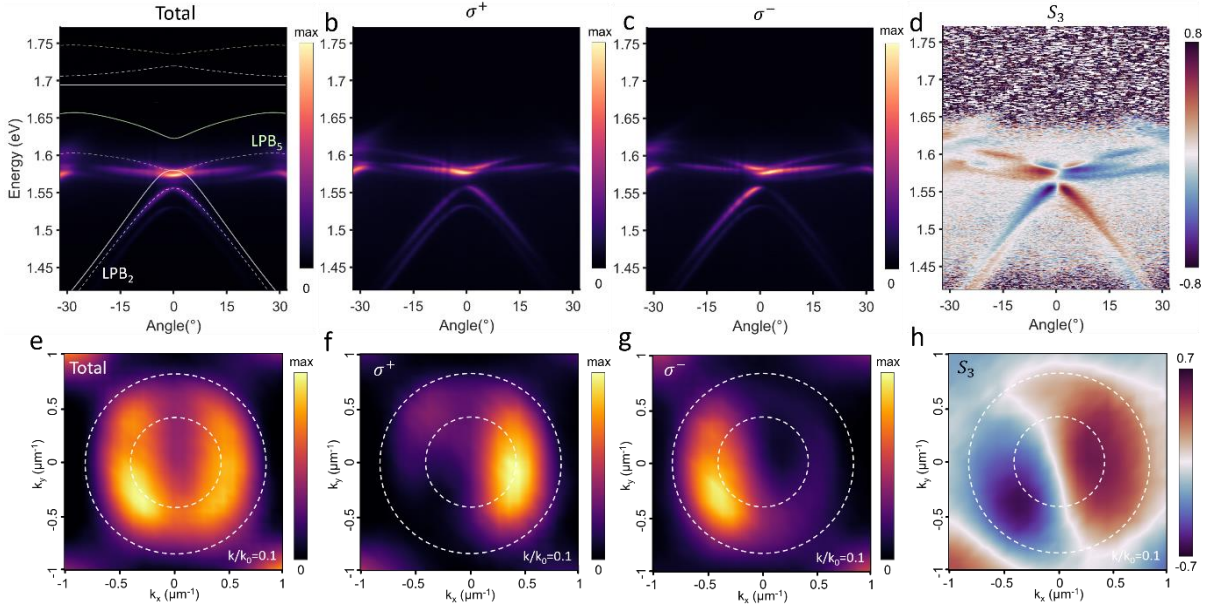

**Figure 3. Electrically driven polaritonic Rashba effect from light-emitting perovskite metasurfaces**. a) Total angle-resolved electroluminescence (EL) spectra for the perovskite metasurface under the positive bias condition with $V_s$ = 0 V, $V_d$ = 90 V and $|V_g|$ = 90 V, showing symmetrically distributed polaritonic states around the normal direction ($k_x$ = 0). The solid lines denote the dispersion of the excitonic and photonic modes while the dashed lines represent polaritonic bands, as defined in Fig. 2f. b, c) The right-circularly ($\sigma^+$) and left-circularly ($\sigma^-$) polarized angle-resolved EL spectra, show mirror symmetric distributions with respect $k_x$ = 0. d) The $S_3$ colormap of the EL dispersion reveals highly pure, spin-polarized bands, a clear manifestation of the polaritonic Rashba effect. e, f, g) Two-dimensional (2D) momentum space images of the total, right-circularly polarized ($\sigma^+$) and left-circularly polarized ($\sigma^-$) polaritonic EL, obtained at the energy of 1.55 eV with a 10-nm band-pass filter. h) The corresponding $S_3$ colormap in the 2D k-space.

**Electrical control of spin-polarized exciton polaritons**

The unique device functionality of the light-emitting metatransistor allows to select momentum and spin of exciton polaritons by simply adjusting the source-drain biasing conditions. While under positive bias condition ($V_d$ > 0, Fig. 1b) the distribution of excitons within the channel is uniform, under the S-biased condition ($V_s$ = -90 V, $V_d$ = 0 V and $|V_g|$ = 80 V, Fig. 1d) excitons are predominantly concentrated near the drain electrode; conversely, under the D-biased condition ($V_s$ = 0 V, $V_d$ = -90 V, Fig. 1f), excitons concentrate near the source electrode. This is clearly seen in the optical images of the emitting device in real space, where distinct EL gradients across the channel can be observed in S- and D-biased conditions (insets of Fig. 4a and 4b and Fig. S9). The corresponding charge injection mechanisms are discussed in Supplementary Note 5 and illustrated in Fig. S10.

Asymmetric exciton distributions in the transistor channel result in the excitation of photonic modes with specific in-plane group velocity, similarly to what had been shown in simulations reported in Ref. 34 (Fig. S7). Experimentally, polariton modes with defined group velocity were demonstrated upon injection of propagating polaritons through the boundaries of a BIC metasurface.[10] In our device, polaritons are generated preferentially on the drain or source side of the metasurface upon electrical injection. When excitons accumulate on the left side of the metasurface, only photonic modes with $u_x > 0$ are populated, forming polaritons that propagate in the $+x$ direction and decay as photons carrying the corresponding energy and momentum. Likewise, when excitons are formed at the right side of the metasurface, the emission derives primarily from polaritons with negative in-plane group velocity $u_x < 0$. It has to be noted that the polariton group velocity is linked to its spin state. Since the metasurface has broken IS but maintains translational symmetry, both the chirality and the group velocity of the polaritonic bands are anti-symmetric with respect to the $k_y$-$k_z$ plane ($k_x$=0), i.e., $S_3(-k_x) = -S_3(k_x)$ and $u(-k_x) = -u(k_x)$. Thus, as the group velocities associated with each pair of circularly polarized states spawning from the BIC are characterized by opposite signs ($u_{\sigma^+} = -u_{\sigma^-}$), the spin states of the polaritons can be selected by controlling the sign of the group velocity.

Figs. 4a and 4b show the angle-resolved electroluminescence (EL) spectra of the device under S-biased and D-biased conditions, demonstrating efficient population of either the right or the left halves of the LPB$_2$ band. To depict the spin state of the excited polaritons on the LPB$_2$ band, the angle-dependent circularly polarized EL intensity at the energy of 1.545 eV, under the two biasing conditions, are shown in Figs. 4c and 4d. Unlike the positive bias case, characterised by equally populated spin-polarized polariton bands (Fig. 3), in the S-biased device we detect strong chiral EL with $S_3$=0.67 primarily from the $\sigma^+$ polaritons at +2º, while the EL of the $\sigma^-$ state is significantly suppressed (Fig. 4c). On the other hand, enhanced chiral EL with $S_3$= -0.54 from the $\sigma^-$ polaritons at -2º is observed in the D-biased device (Fig. 4d). This demonstrates electrical selection of spin states and directionality of the exciton polaritons.

The strong enhancement of the spin-polarized polaritonic emission at ±2º, compared to the excitonic emission of the bare perovskite film (shaded areas in Figs. 4c and 4d), is due to the high Q-factor near the C-points, discussed in Fig. 2d. It shall be noted that it is possible to tune the chiral EL emission angle by adjusting the polariton energy away from the C-points, albeit at the cost of sacrificing spin purity (a reduction of $S_3$), as shown in Fig. S11.

In addition to the electrical selection of polariton spin states that emit in opposite directions from a single metasurface, the design of pixelated metasurfaces can be employed to implement electrical switching between polaritons with opposite spins emitting in a given direction. For instance, this is possible using two metasurfaces with 180º-rotated triangular holes positioned next to the source and drain electrodes (Fig. S12). Under S-biased or D-biased conditions, the emission from each pixel can be assumed to be uniform. When transitioning between the two pixels by switching from the S-biased to the D-biased condition, the EL emission intensity in a given direction remains nearly unchanged, but its polarization is flipped (Fig. S12). This provides an alternative route for spin polariton control when fixed emission directionality is required, such as to realize polarized light sources for quantum information processing.

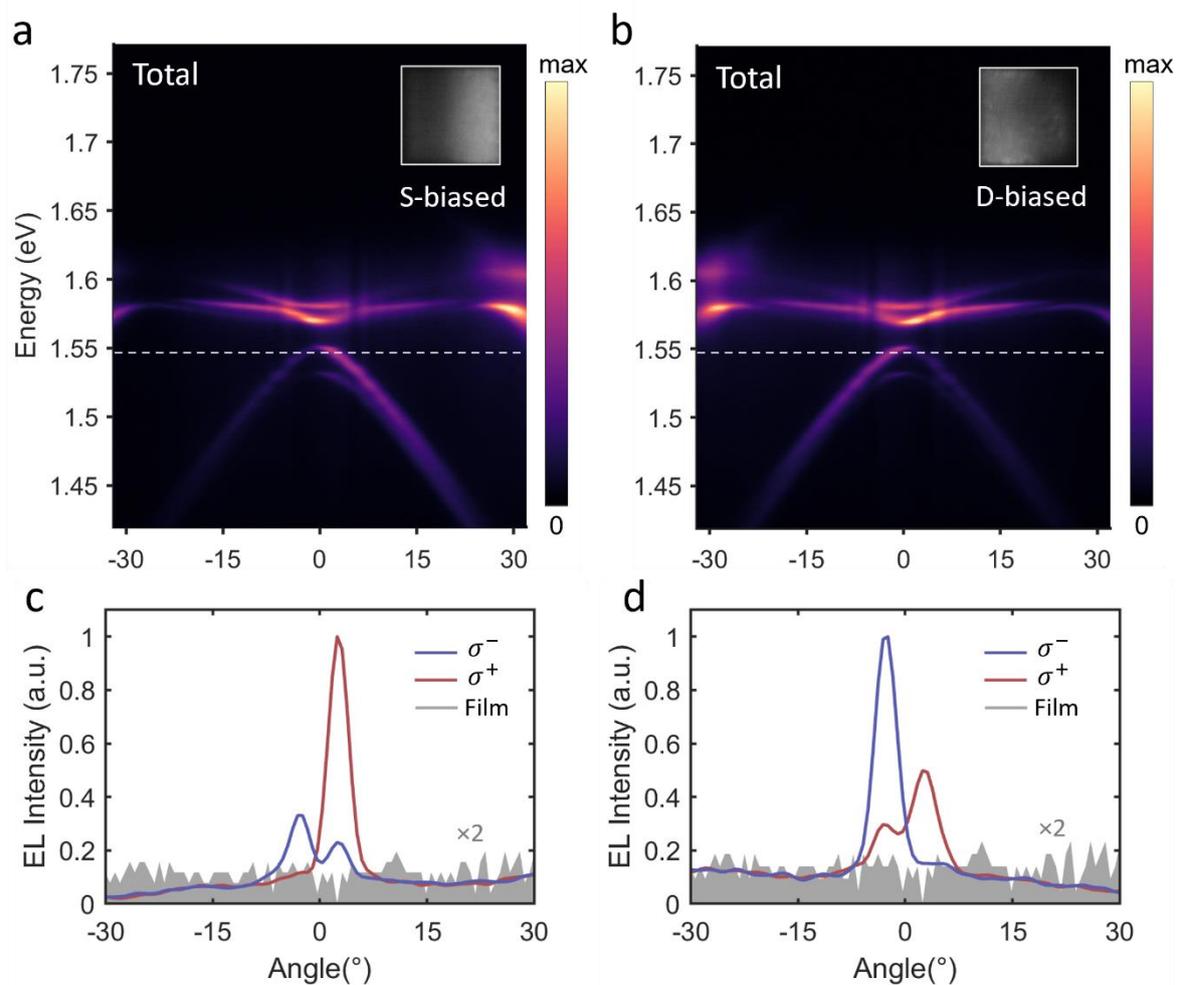

**Figure 4. Electrically tunable polariton spins and directionality from light-emitting perovskite metasurfaces.** a, b) The total angle-resolved electroluminescence spectra, under D-biased and S-biased conditions, show a remarkable resemblance to the right- and left-circularly polarized EL spectra of Fig. 3b,c, thus demonstrating selective population of the spin-polariton modes. The insets show the real

images of the EL emission from the perovskite metasurface, under the respective bias condition. c, d) Spin-polarized polaritonic EL intensity at 1.545 eV (white dashed lines in Figs. 4a,b), exhibiting a strong degree of circular polarization at ±2º. The grey shaded areas indicate the angle dependent EL intensity of the unpatterned perovskite film, at the same photon energy.

**Discussion**

In conclusion, our study has successfully demonstrated the generation of directional spin-polarized exciton polaritons from monolithic dielectric metasurfaces upon electrical injection, a feat previously accomplished only in optically pumped microcavities. The monolithic integration offers significant advantages by enhancing light-matter interaction and reducing the overall complexity of the system. In our work, the broken in-plane IS of the metasurface lifts the spin degeneracy of the photonic modes, while the broken out-of-plane IS enhances the Q-factor of the spin states. Strong coupling is achieved between the electrically injected perovskite excitons and the cavity photons at 78 K, with the Rabi splitting energy of ~120 meV and high purity of directional spin-polarized polaritons ($|S_3| = 0.8$).

Leveraging on spin-momentum locking, the unique functionality of the light-emitting transistor allows asymmetric exciton injection under different biasing conditions, providing the capability to electrically control the spin state and directionality of the polaritonic emission. This platform could be readily extended to implement electrical tunability of other characteristics of exciton polariton emission, e.g., amplitude and phase, simply by varying the metasurface design. Thus, we believe this accomplishment takes a significant step toward the integration of polariton spintronic technologies with electronics, facilitating the practical realization of polaritonic devices and architectures such as electrically driven and tunable inversionless spin lasers, or spin logic circuits.

**Methods**

**Numerical simulations**

The eigenmode simulations for the photonic band structure, Q-factor, and normalized Stokes parameter were conducted employing COMSOL Multiphysics 5.4. Floquet periodic boundary conditions were applied in the transverse direction, while perfectly matched layers were utilized in the z-direction. The Stokes parameter was determined by simulating the far-field Jones vector $[\widetilde{E_x}, \widetilde{E_y}]^T$ of the eigenmode. Angle-resolved reflection spectra were simulated using rigorous coupled wave analysis (RCWA) within the Ansys Lumerical software. The refractive index of the MAPbI$_3$ perovskite at 78 K was derived from tabulated permittivity data in ref. 48, fitting the data with a Tauc-Lorentz model. This methodology facilitated the extraction of optical constants with and without excitonic resonance. Further details are provided in Supplementary Note 1.

**Material synthesis**

MAPbI$_3$ thin films were synthesized using a solution-processed spin-coating method. Methylammonium iodide (CH$_3$NH$_3$I, MAI) was procured from Greatcell, lead iodide (PbI$_2$, 99.99%) from Tokyo Chemical Industry Co. Ltd., and anhydrous dimethylformamide (DMF) from Sigma Aldrich. All chemicals were used as received without purification. The 1.2 M MAPbI3 precursor solution was prepared by dissolving MAI and PbI$_2$ powder in DMF with a molar ratio of 1:1. After magnetic stirring for 2 hours at 273 K in a N2-filled glovebox, the solution was filtered through a polyvinylidene fluoride (PVDF) syringe filter (0.45 μm) before spin-coating. Substrates, heavily p-doped Si with a thermally grown SiO$_2$ (500 nm) layer, were cleaned with ultrasonication in acetone, isopropanol, and deionized water. Following drying and an oxygen plasma cleaning treatment, the perovskite precursor solution was spin-coated onto the substrates with a speed of 4900 rpm for 35 s using the anti-solvent deposition method, with toluene drop-cast on the substrates 5 s after starting spinning. The resulting films were finally annealed at 373 K for 15 min, yielding a 350 nm thickness film, as indicated by atomic force microscopy (AFM) in Fig. S13.

**Light emitting device and metasurface fabrication**

The perovskite light emitting device adopts the similar lateral configuration of a thin film field effect transistor using a bottom-gate and top-contact configuration. After the perovskite thin film fabrication, 100 nm thick Au electrodes were thermally evaporated in high vacuum (~10$^{-6}$ mbar) using a shadow mask. To avoid thermal decomposition of the perovskite films, samples were placed on a water-cooled substrate holder at 291 K during the electrode deposition. The resulting LET channel length (L) and width (W) were 80 μm and 1 mm, respectively. The perovskite metasurface fabrication was conducted by focused ion beam (FIB) lithography process. A 40 μm × 40 μm triangular air hole array was patterned between the source and drain electrodes using the Helios 600 NanoLab, FEI system. The ion beam current was controlled around ~4 pA for suitable spot size of etching.

**Electrical and optical characterization**

Electrical and optical characterization was carried out using a temperature-controlled probe stage (HFS600E-PB4/PB2, Linkam) at the temperature of 78 K in the dark and under vacuum (~$10^{-3}$ mbar). DC measurements for the transfer characteristics were acquired with a 2-channel precision source/measure unit (B2902A, Agilent). Charge-carrier mobilities were extracted from the forward sweeping of transfer characteristics obtained at $V_{ds}$ = 80 V, using the conventional equations for metal-oxide semiconductor (MOS) transistors in the saturation regime: $\mu_{sat} = \frac{2L}{WC_i}\left(\frac{\partial \sqrt{I_{ds}}}{\partial V_g}\right)^2$.

The PL and EL measurements were performed with a home-built micro-spectrometer. The system consists of an inverted optical microscope (Nikon Ti-U), a spectrograph (Andor SR-303i with 150 lines/mm grating), and an electron-multiplying charged-coupled detector (EMCCD, Andor Newton 971). The PL spectra of the perovskite film were taken when the sample was excited by a 405 nm continuous wave solid state laser with a spot size of ~15 μm in diameter. The EL measurements of unpatterned light emitting devices were performed on the same optical setup under positive bias mode of $V_s$ = 0V, $V_d$ = 90V and AC gate voltage of $|V_g|$ = 80V (100 kHz modulation frequency), using an arbitrary waveform generator (3390, Keithley) coupled with a high-voltage amplifier (WMA-300, Falco Systems). The optical images of the operating device were taken and acquired by a cooled sCMOS scientific camera (PCO.edge 3.1m) coupled to the optical microscope.

**Angle-resolved measurements**

The angle resolved reflection, PL, and EL measurements of the perovskite metasurface were also performed in the home-built micro-spectrometer setup. A lens system along the light path between the microscope and the spectrograph is used to project the back focal plane of the collection objective (Nikon ×50, with numerical aperture NA=0.45) onto the slit of the spectrograph. This configuration allows spectral or spatial measurement with angular information corresponding to the NA of the objective. A pinhole in the detection path is used as a spatial filter to collect the signal only from the center of the metasurface, which mitigates the edge effects. A linear polarizer and quarter waveplate placed on the optical path of the lens system were used to select the collection polarization. A halogen white light was used as the light source for the reflection measurement. The angle-resolved PL and EL spectra were taken using the same pumping scheme as the previous section. The EL spectra with S-biased and D-biased conditions were conducted in the same setup by changing the electrical bias on the source and drain electrodes.

## Acknowledgements

Research was supported by the Singapore Ministry of Education (Grant no. MOE-T2EP50222-0015) and by the Agency for Science, Technology and Research, A*STAR (A*STAR-AME Grant no. A18A7b0058). S.T.H acknowledges support from A*STAR MTC-Programmatic (Grant No. M21J9b0085).

## Author contributions

C.S., J.T. G.A. and Y.W. conceived the idea and planned the research. Y.W. fabricated the perovskite devices, performed the electrical and optical characterization of the sample, and conducted the theoretical analysis, numerical simulations and data processing. G.A. fabricated the metasurfaces. S.T.H. and J.T. contributed to optical characterization and J.T. helped with numerical simulations. Y.W, J.T., G.A. and C.S. performed data analysis and wrote the paper with inputs from S.T.H.. C.S. and G.A. supervised the work.

## Data availability

The data that support the findings of this study are openly available in the NTU research data repository DR-NTU (Data) at

## Conflict of interest

The authors declare no competing interests.

Supplementary Information for

# Electrically generated exciton polaritons with spin on-demand

Yutao Wang,[1,2,3] Giorgio Adamo,[1,2] Son Tung Ha,[4] Jingyi Tian,[1,2] and Cesare Soci[1,2*]

[1] *Centre for Disruptive Photonic Technologies, TPI, Nanyang Technological University, 21 Nanyang Link, Singapore 637371*

[2] *Division of Physics and Applied Physics, School of Physical and Mathematical Sciences, Nanyang Technological University, 21 Nanyang Link, Singapore 637371*

[3] *Interdisciplinary Graduate School, Energy Research Institute @NTU (ERI@N), Nanyang Technological University, 50 Nanyang Drive, Singapore 637553*

[4] *Institute of Materials Research and Engineering, Agency for Science Technology and Research (A*STAR), 2 Fusionopolis Way, Singapore 138634*

*\*Correspondence: csoci@ntu.edu.sg*

## Table of Content:

**Supplementary Note 1: Optical characterization of MAPbI3 perovskite thin film**

To obtain the excitonic resonance energy and damping rate of the MAPbI3 perovskite thin film, a spectroscopic analysis was undertaken on the near-gap absorption features at 78 K. The steady-state absorption spectra of the perovskite thin film are illustrated by the black line in Figure S1. The disentanglement of the excitonic and continuum contributions was accomplished through the application of Elliott's formula[1], expressed as follows:

$$Abs(E) = AE^{-1}\left(E_b^{\frac{3}{2}}\sum_n^{\infty}\frac{4\pi}{n^3}\delta(E-E_{exc}) + \frac{2\pi\sqrt{\frac{E_b}{E-E_g}}}{1-\exp\left(-2\sqrt{\frac{E_b}{E-E_g}}\right)} \cdot B \cdot \begin{cases}\sqrt{E-E_g} & if\ E > E_g \\ 0 & if\ E > E_g\end{cases}\right),$$

where E is the energy of the photon, A and B are constants, $E_b$ is the exciton binding energy and $E_g$ is the bandgap. The first term describes the contribution of a series of excitons below the bandgap and the second term indicates the contribution from the continuum spectra. In order to fit the absorption spectra with this formula, the unrealistic delta function and the step functions are mimicked by Gaussian distribution and Boltzmann sigmoidal function with a finite spectral broadening. The equation can be rewritten as the following:

$$Abs(E) = \left(C\cdot 4\pi E_b^{\frac{3}{2}}\frac{1}{\sqrt{2\pi}w_1}\exp\left(-\frac{(E-E_g+E_b)^2}{2\cdot w_1^2}\right) + D\cdot\frac{1}{1+\exp\left(\frac{E_g-E}{w_2}\right)}\right),$$

with C and D constants. The fitted spectra are indicated by the brown solid line in Fig. S1, showing perfect agreement with the experimental result. The fitted parameter and value are indicated in Table S1.

Table S1. Fitted parameters of the Elliot model for the MAPbI3 absorption spectra.

| Parameter | Value |
|---|---|
| $E_b$ | 0.032 eV |
| $E_{exc}$ | 1.693 eV |
| $E_g$ | 1.725 eV |
| $w_1$ | 0.028 eV |
| $C$ | 0.535 eV$^{-1/2}$ |
| $D$ | 0.684 |
| $w_2$ | 0.020 eV |

The exciton energy and the damping rate are extracted to be 1.693 eV and 68 meV, respectively. The exciton binding energy is extracted to be around 32 meV, comparable to the previously reported literature[2].

The red and blue solid line in Fig. S1 indicates the PL and EL spectra of the perovskite film. The spectra show similar line shape with slight red shift from the EL, which may be caused by the re-absorption and re-emission of the EL photons from the bottom of the perovskite layer.

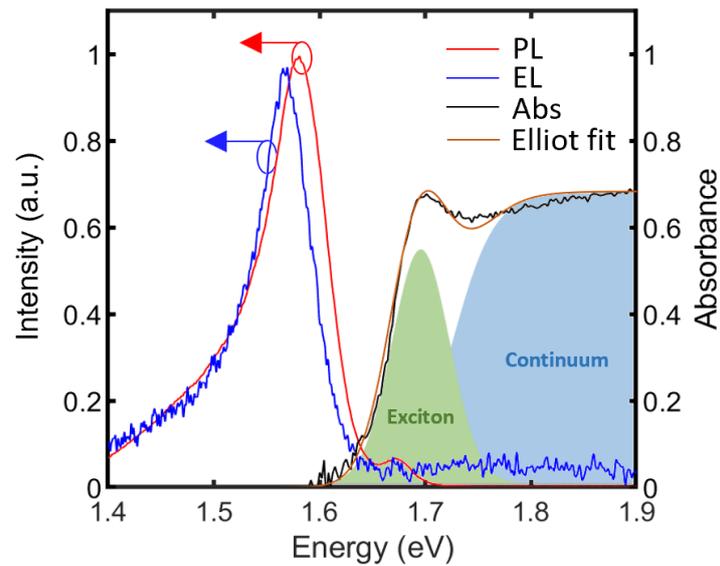

**Figure S1. Absorbance, EL, and PL spectra of the perovskite film at 78 K.** The brown line is the spectra from Elliot fit model. The exciton and continuum spectra are indicated as the green and blue shaded area, respectively.

**Supplementary Note 2: Electrical characteristic of MAPbI$_3$ perovskite transistor**

We characterized the electrical curve of the device at 78 K to study the mobility of the charge carriers. Fig. S2a and b shows the p and n-type transfer curve of the transistor under different source-drain bias ($V_{ds}$), showing high on-off ratio of $10^3$ ($10^2$) for n (p) type device. The output characteristics in Fig. S2c and d shows strong gate modulation and distinguishable linear and saturation regime. The forward (solid lines) and backward (dashed lines) scanning undergoes a hysteresis behaviour because of the ionic motion in the perovskite layer. The charge-carrier mobilities were extracted from the forward sweeping of transfer characteristics obtained at $V_{ds}$ = 80 V, using the conventional equations for metal-oxide semiconductor (MOS) transistors in the saturation regime: $\mu_{sat} = \frac{2L}{WC_i}\left(\frac{\partial \sqrt{I_{ds}}}{\partial V_g}\right)^2$, yielding $\mu_e = 1.2 \times 10^{-2}\ cm^2 \cdot V^{-1} \cdot s^{-1}$ and $\mu_h = 1.0 \times 10^{-3}\ cm^2 \cdot V^{-1} \cdot s^{-1}$. The unbalanced electron and hole mobility plays an important role of electrically tunable emission zone in the light emitting transistor channel, which will be further discussed in Supplementary Note 4.

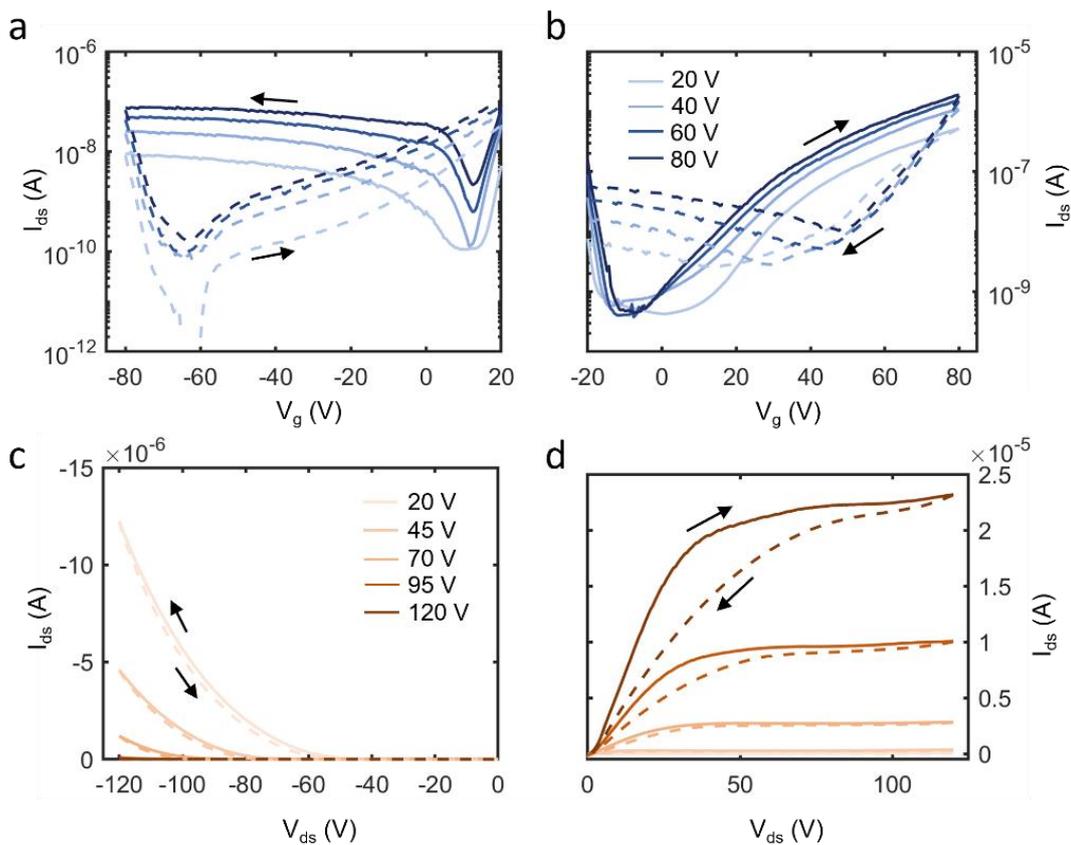

**Figure S2. Electrical characterization of MAPbI3 perovskite transistor at 78 K.** a, b) p-type and n-type transfer characteristics of the perovskite transistor respectively with $V_{ds}$

varying from 20 V to 80 V. c, d) p-type and n-type output characteristics of the perovskite transistor, respectively, with $V_g$ varying from 20 V to 120 V. The electron and hole mobility were extracted to be $\mu_e = 1.2 \times 10^{-2}\ cm^2 \cdot V^{-1} \cdot s^{-1}$ and $\mu_h = 1.0 \times 10^{-3}\ cm^2 \cdot V^{-1} \cdot s^{-1}$ from the saturation regime.

**Supplementary Note 3: Tauc-Lorentz material model for MAPbI₃ perovskite**

The Tauc-Lorentz model describes the optical response of the dielectric material using the Tauc joint density of states and the Lorentz oscillator, which gives the complex dielectric function:

$$\widetilde{\varepsilon_{TL}} = \varepsilon_{r,TL} + i \cdot \varepsilon_{i,TL} = \varepsilon_{r,TL} + i \cdot (\varepsilon_{i,T} \times \varepsilon_{i,L})$$

Where the imaginary part $\varepsilon_{i,TL}$ is given by the product of the imaginary part of the Tauc's dielectric function $\varepsilon_{i,T}$ and the Lorentz one $\varepsilon_{i,L}$.

As a result, the imaginary part of the permittivity according to the Tauc-Lorentz model is given by:

$$\varepsilon_{i,TL} = \begin{cases} \sum_{k=1}^{n} \frac{1}{E} \times \frac{A_k \cdot E_k \cdot C_k \cdot (E - E_g)^2}{(E^2 - E_k^2)^2 + C_i^2 \cdot E^2}, & for\ E \geq E_g \\ 0, & for\ E < E_g \end{cases}$$

Where $A_k$, $E_k$ and $C_k$ are the amplitude, central energy and the damping rate of the k-th oscillator. $E$ is the photon energy and $E_g$ is the optical band gap.

The real part of the permittivity is retrieved from the imaginary part by the Kramers-Kronig (K-K) relations:

$$\varepsilon_{r,TL} = \varepsilon_r(\infty) + \sum_{k=1}^{n} \frac{2}{\pi} \cdot P \cdot \int_{E_g}^{\infty} \frac{\xi \cdot \varepsilon_k(\xi)}{\xi^2 - E^2} d\xi$$

Where $\varepsilon_r(\infty)$ is the background permittivity and it was fitted to be 2.16.

**Table S2. Fitted parameters for the MAPbI₃ Tauc-Lorentz dielectric at 78 K.**

| k | $A_k$ | $E_k$ | $C_k$ |
|---|---|---|---|
| 1 (exciton) | 18.059 | 1.675 | 0.035 |
| 2 | 27.088 | 1.671 | 0.150 |
| 3 | 3.386 | 1.851 | 0.249 |
| 4 | 20.316 | 2.430 | 0.597 |
| 5 | 9.029 | 3.798 | 1.002 |

In order to acquire the refractive index dispersion of the passive material without the exciton resonance. We analyze the permittivity dispersion of the MAPbI₃ perovskite as a Tauc-Lorentz dielectric. The permittivity of the MAPbI$_3$ perovskite was acquired from the tabulated data[3]. We fit the Tauc-Lorentz model to the data to obtain the parameters for the oscillators and the fitted results is shown in Table. S2. The 1$^{st}$ oscillator with the smallest damping rate is the excitonic resonance.

Fig. S3a shows the fitted dispersion result of the imaginary part of the permittivity using the Tauc-Lorentz model. The retrieved real part is shown in Fig. S3b, both showing good agreement with the acquired data. The corresponding imaginary and real part of the refractive index are shown as the black lines in Fig. S3c and d, respectively. After excluding the excitonic oscillator, we can acquire the dispersion relation of the passive refractive index, indicated as the orange dashed lines in Fig. S3c (imaginary part) and Fig. S3d (real part).

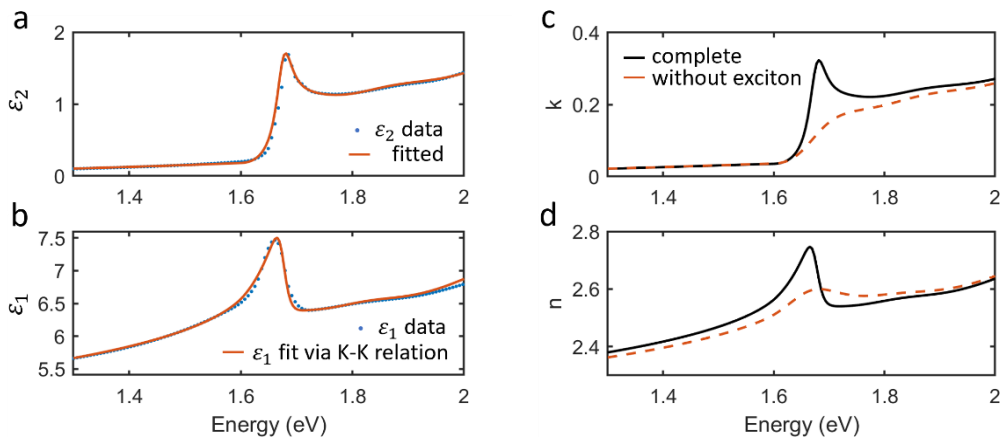

**Figure S3. Tauc-Lorentz fits of MAPbI$_3$ perovskite film.** a) Imaginary part of the MAPbI$_3$ perovskite permittivity at 78 K. Blue dots indicates the data obtained from ref [3]. The orange line indicates the Tauc-Lorentz model fit. b) Real part of the MAPbI$_3$ perovskite permittivity at 78 K. Blue dots indicates the tabulated data and the orange line is the $\varepsilon_1$ fit calculated from $\varepsilon_2$ via the K-K relation. c) Imaginary part and d) real part of the refractive index with (black line) and without exciton resonance (orange dashed line).

**Supplementary Note 4: MAPbI$_3$ metasurface with broken out-of-plane IS**

The broken in-plane IS in the triangular metasurface introduced the lifted spin degeneracy in the photonic band. However, the Q-factor of the circularly polarized state is relatively low as shown in Fig. 2d in the main text, which hinders the application in strong light matter interaction. Fig. S4a shows the photonic bands structure without breaking the out-of-plane IS of the system, showing that the Ph$_1$ (TE-like) and Ph$_2$ (TM like) band are nearly degenerate as the k$_x$ increases. Upon breaking the out-of-plane IS of the system by introducing a 100 nm residual layer and the SiO$_2$ substrate, as indicated in Fig. S4b, the Ph$_1$ and Ph$_2$ band no longer cross with each other due to the TE-TM hybridization. Simpler to the generation of Friedrich-Wintegen BIC via vertical symmetry breaking, the interaction of Ph$_1$ and Ph$_2$ bands will result in the formation of one new eigenstates with lower losses[4], which could lead to the enhancement of the Q-factor of the C-point photonic state. It is also noted that the TE-TM hybridization is still remained while adding additional Si layer below the SiO$_2$ dielectric, as indicated in Fig. S4c. The Q-factor distribution of the Ph$_2$ band in the three cases are simulated in Fig. S4d to f. A clear peak Q-factor off the Γ-point can be observed in the latter two cases with broken out-of-plane symmetry.

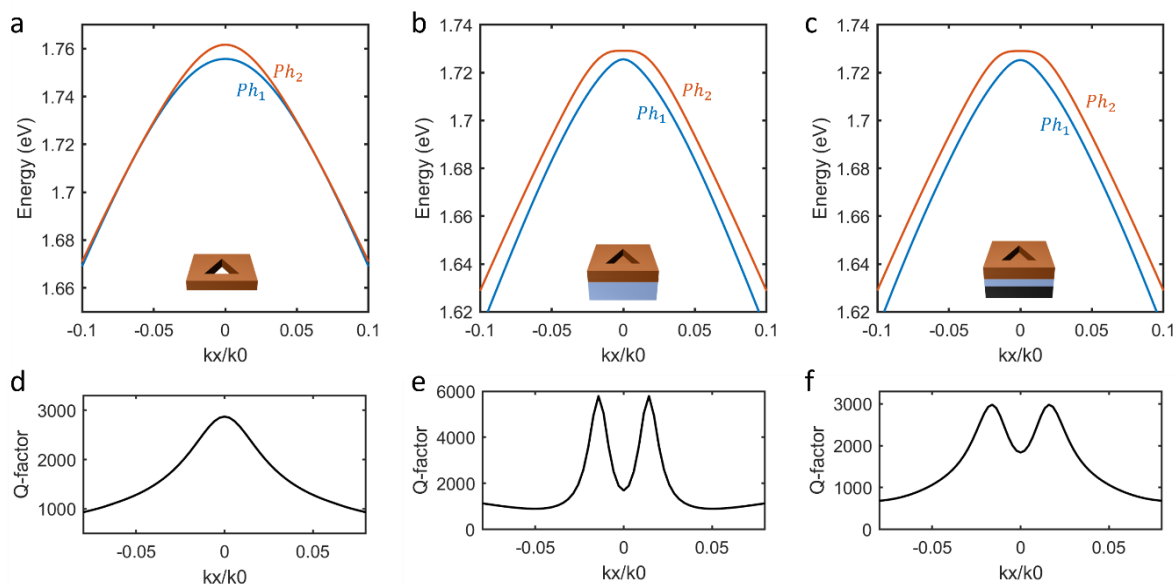

**Figure. S4 Band structure and corresponding Q-factor of the metasurface.** Diagram of the Ph$_1$ and Ph$_2$ photonic band structure in the system with a) perfect out-of-plane IS, b) broken out-of-plane IS with residual layer and SiO$_2$ substrate and c) Si substrate with 500 nm SiO$_2$ layer. The TE-TM mode coupling between the Ph$_1$ and Ph$_2$ band arises from the broken out-of-plane symmetry, which results in the anti-crossing behavior of the two modes in the latter two cases. d)-f) Q-factor distribution of the Ph$_2$ photonic band in the system shown in panel a)-c).

We systematically examined the Q-factor and distribution of S$_3$ within the Ph$_2$ band through the manipulation of the SiO$_2$ layer thickness to identify an optimal device design. Given that reducing the dielectric layer thickness may lead to elevated leakage current, while excessively thick layers may impede the gating effect, both adversely impacting device performance, we constrained the considered thickness range to be between 100 and 700 nm. The device with a thickness of 100 nm is omitted from consideration due to its generally low Q-factor, as depicted in Figure S5a. Despite the observation of an augmented Q-factor at the C-point in all other three designs, the selection of a 500 nm SiO$_2$ thickness is ultimately made, as it sustains a high level of S$_3$ across a broader angular range.

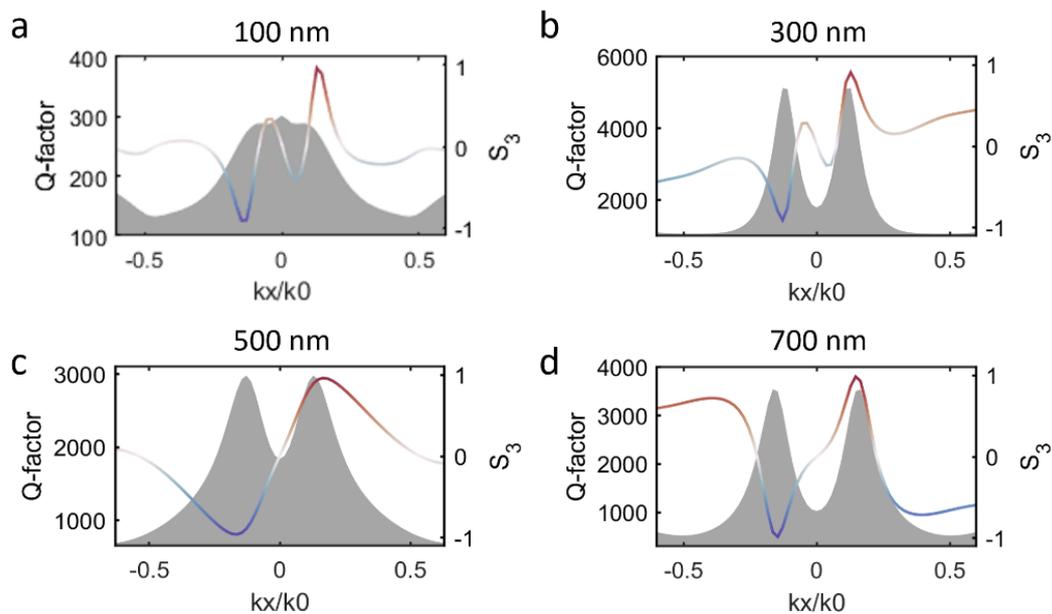

**Figure. S5 Q-factor and S$_3$ distribution with different SiO$_2$ thickness.** Simulated Q-factor distribution and the S$_3$ parameter in the Ph$_2$ band with SiO$_2$ thickness of a) 100 nm, b) 300nm, c) 500 nm and d) 700 nm between the perovskite and Si layer. The 500 nm design is chosen in our work for experimental demonstration.

Apart from the afore mentioned eigenmode simulation, we also conducted rigorous coupled wave analysis (RCWA) to simulate the Q-factor distribution in the Ph$_2$ band. As shown in Fig. S6a, a peak Q-factor at the angle of ~2.4° can be found, further confirms the high Q-factor at the C-point upon out-of-plane symmetry breaking. Moreover, the experimental Q-factor distribution is extracted from the angle-resolved EL measurement, confirming a peak Q-factor off the Γ-point.

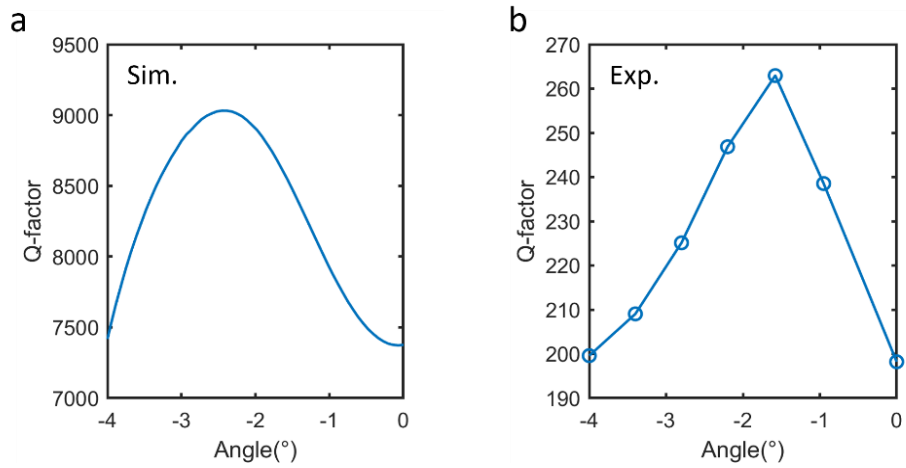

**Figure. S6 Simulated and experimental Q-factor distribution.** a) Simulated and b) experimental Q-factor of the $LPB_2$ near the Γ-point. The simulated Q-factor from the RCWA simulation is derived using a Fano fit on the reflection spectra and the experimental one is derived from the EL spectra using a Lorentzian fit. Clear peak Q-factor off the Γ-point can be observed in both cases.

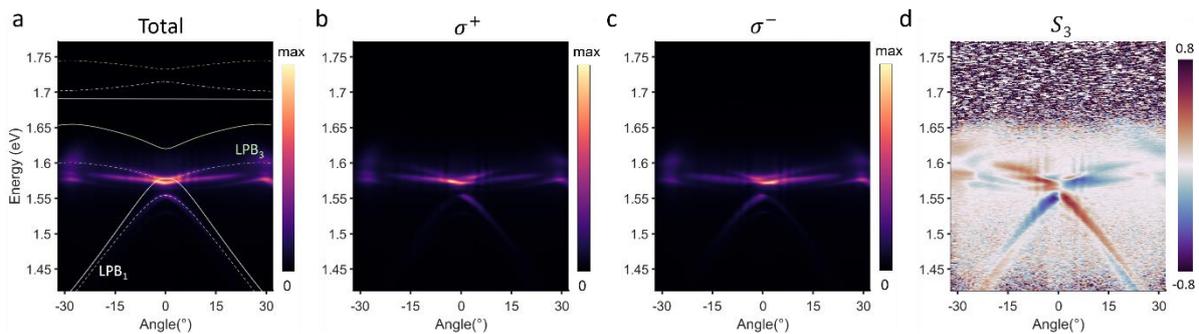

**Figure S7. Polariton Rashba effect under optical excitation.** a) Total angle resolved PL spetra showing massive occupation of the polaritons in the lower polariton bands. The solid and dashed lines in the figure denotes the exciton, photon and polariton bands as in Fig. 3a. b), c) The right-circularly ($\sigma^+$) and left-circularly ($\sigma^-$) polarized angle-resolved PLL spectra. d) Corresponding $S_3$ colour map showing splitted polariton spin in the opposite half of the k-space.

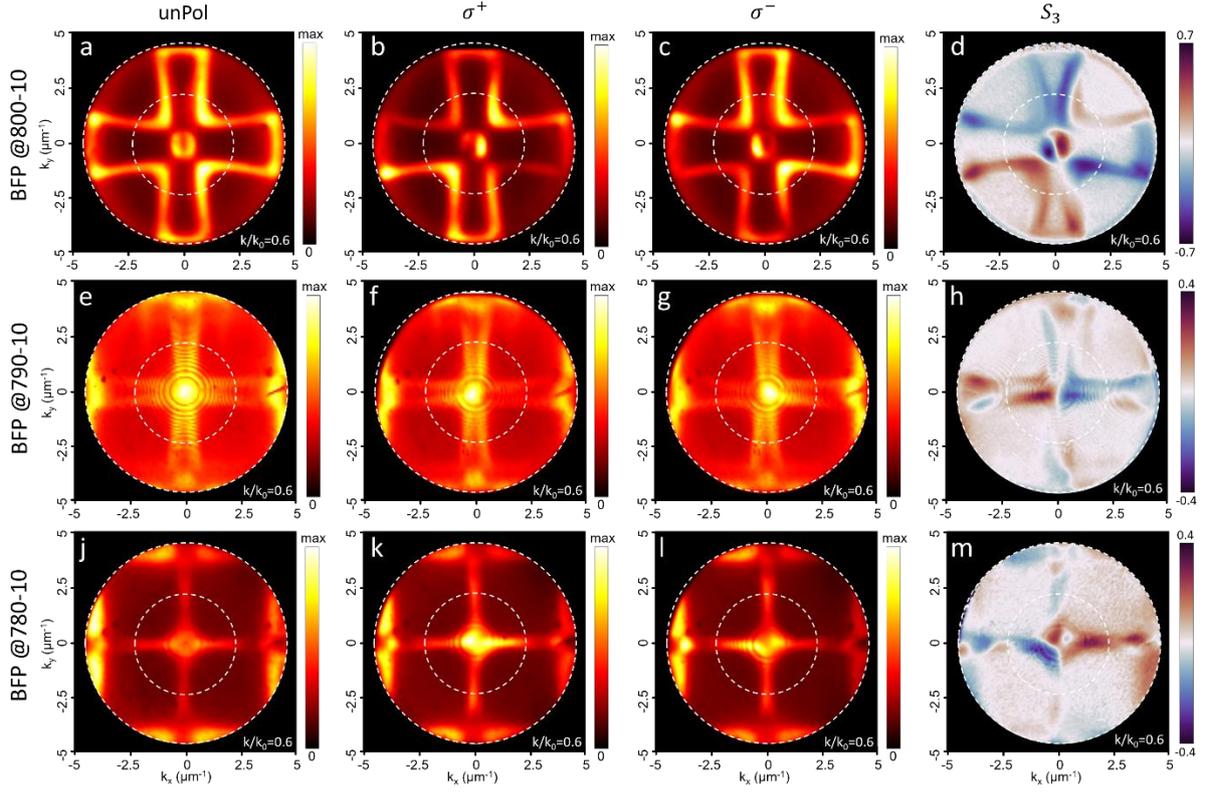

**Figure S8. Back focal plane (BFP) image of spin-polarized polariton electroluminescence.** a-d) Unpolarized, $\sigma^+$ $\sigma^-$ polariton electroluminescence and the corresponding $S_3$ value at the energy of 1.55 eV, taken with a 800 nm bandpass filter (bandwidth of 10 nm). e-f) Same measurement as a-d at the energy of 1.57 eV, taken with a 790-10 nm bandpass filter. j-m) Same measurement taken at the energy of 1.59 eV, taken with a 780-10 nm bandpass filter.

**Supplementary Note 5: Electrically tunable exciton distribution**

The exciton distribution within the transistor channel can be electrically tuned in the x direction. Fig. S9 shows the optical image of the operating light emitting device under different biasing condition. In the symmetric biased case, relatively uniform and enhanced emission from the metasurface area can be observed. However, in the D-biased and S-Biased case, the emission majorly comes from the region close to the source or the drain electrode, implying a non-uniform exciton distribution in the x direction.

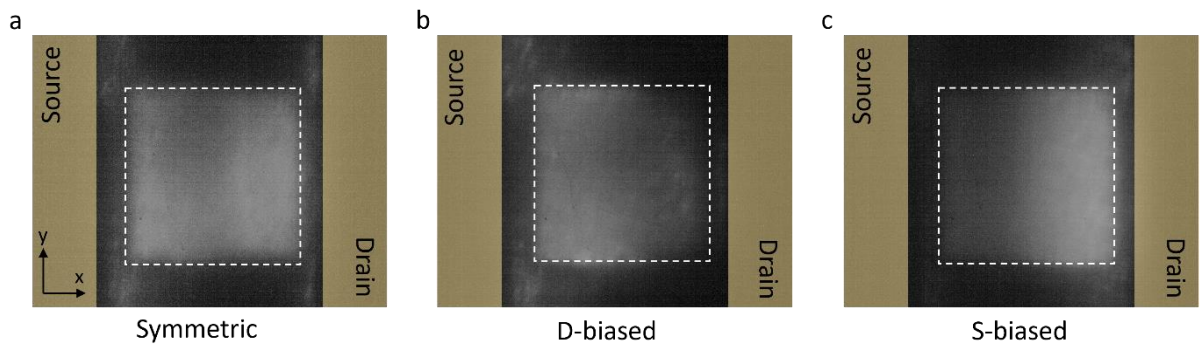

**Figure S9. Optical microscopy image of the operating light emitting device with different exciton distributions.** a) Symmetric exciton distribution with the voltage of $V_s = 0$ V, $V_d = 90$ V and $|V_g| = 80$ V. The white dashed square indicates the location of the fabricated metasurface between the source and drain electrodes. Relatively uniform exciton distribution in the x direction can be found in this pumping scheme. b) Schematic of exciton distribution close to the source electrode with $V_s = 0$ V, $V_d = -90$ V and $|V_g| = 80$ V. c) Exciton distribution close to the drain electrode with $V_s = -90$ V, $V_d = 0$ V and $|V_g| = 80$ V.

Schematics of charge carrier injection and exciton formation/recombination within the channel of the light emitting device under different operating biasing conditions are shown in Fig. S10. In the symmetric distribution case (Fig. S10a), electrons and holes are periodically injected (under AC gate bias) from the source electrode to the gate channel above the SiO$_2$ layer; however, since $V_d$ is always larger than $V_g$, only holes are injected from the drain electrode. Because of their higher mobility, electrons can drift through the channel and form excitons with holes injected from both, source and drain electrodes, uniformly across the channel (orange dashed circle). In the D-biased device (Fig. S10b), electrons and holes are also periodically injected from the source electrode but only electrons are injected from the drain electrode because $V_g$ is larger than $V_d$. In this case, holes injected from the source can only bind with electrons injected from the same electrode because of their limited mobility, and excitons are formed predominantly near the source electrode. In the S-biased device, source and drain biases are reversed compared to the D-biased case, leading to the formation of excitons nearby the drain electrode (Fig. S10c).

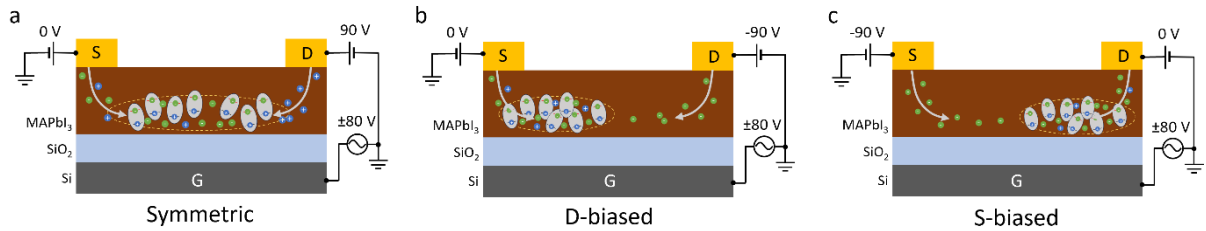

**Figure S10. Principles of electrically tunable exciton distribution in the perovskite film.** a) Symmetric exciton distribution with the voltage of $V_s = 0$ V, $V_d = 90$ V and $|V_g| = 80$ V. The brown, light blue and grey layers indicate the $MAPbI_3$ perovskite, $SiO_2$ and p-doped Si, respectively. The green dots indicate the electrons while the blue dots indicate the holes. The yellow dashed circle indicates the recombination area of the excitons injected from the source and drain electrode. b) Schematic of exciton distribution close to the source electrode with $V_s = 0$ V, $V_d = -90$ V and $|V_g| = 80$ V. c) Exciton distribution close to the drain electrode with $V_s = -90$ V, $V_d = 0$ V and $|V_g| = 80$ V.

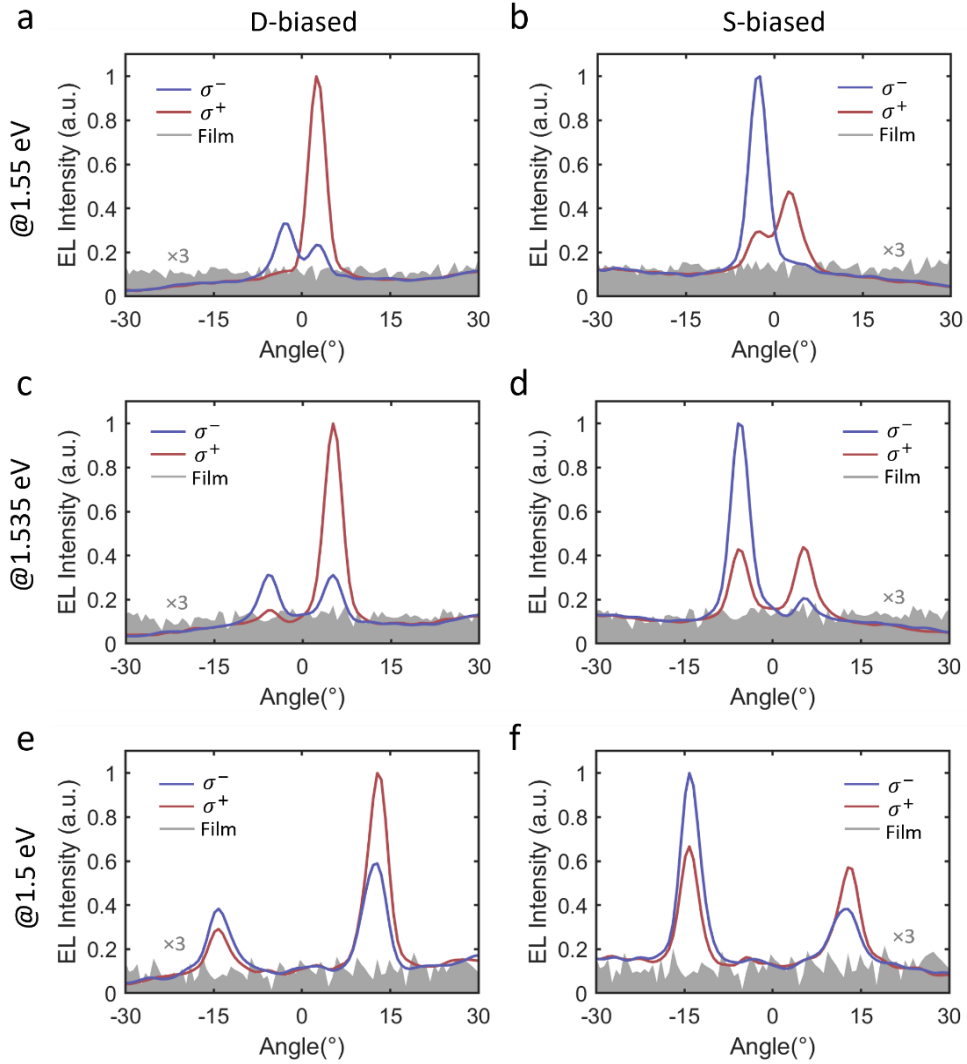

**Figure S11. Electrically tunable spin-polarized polariton EL at different energy.** a), c) and e) Angle-dependent EL of the polaritons in the D-biased device with the energy of 1.55 eV, 1.535 eV and 1.50 eV, respectively. b), d) and f) Same measurement as a, c and e but for the S-biased device. The blue and red lines indicate the emission from the $\sigma^-$ and $\sigma^+$ polaritons, respectively. The grey shaded area indicates the excitonic emission multiplied by 3 from the unpatterned perovskite film at the corresponding photon energy. As the energy reduces in the LPB$_2$ band, the emission angle of EL from the spin-polarized polariton increases to ~ ±14.5° while the spin purity is reduced due to the smaller $S_3$ value in the Ph$_2$ band at large angles.

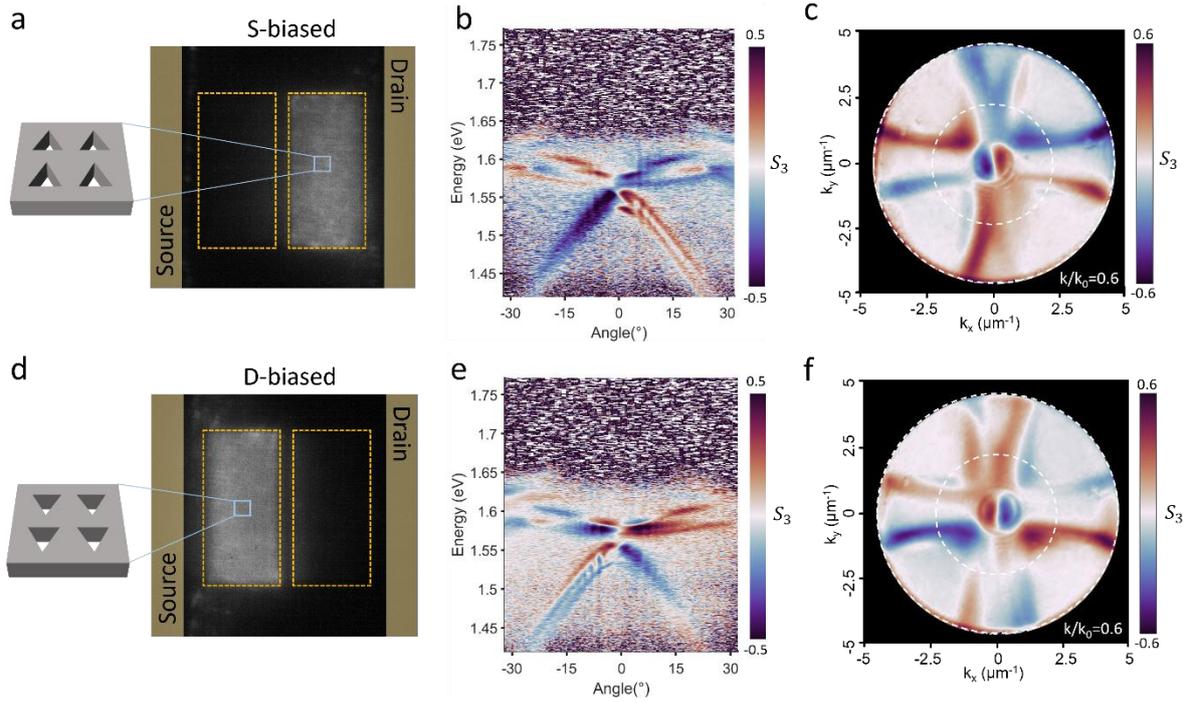

**Figure S12. Electrically tunable polariton spin in pixelated metasurface.** a), c) Optical image of the operating S-biased and D-biased device with a pixelated metasurface design. The dashed rectangle on the right indicates the metasurface with triangular unit cell pointing upward while the dashed rectangle on the left indicates the triangle pointing downward. b), e) Angle-resolved $S_3$ parameter map of the polaritonic EL from the S-biased and D-biased device, respectively. c), f) Back focal plane $S_3$ map of the polariton EL from the S-biased and D-biased device at the energy of 1.55 eV.

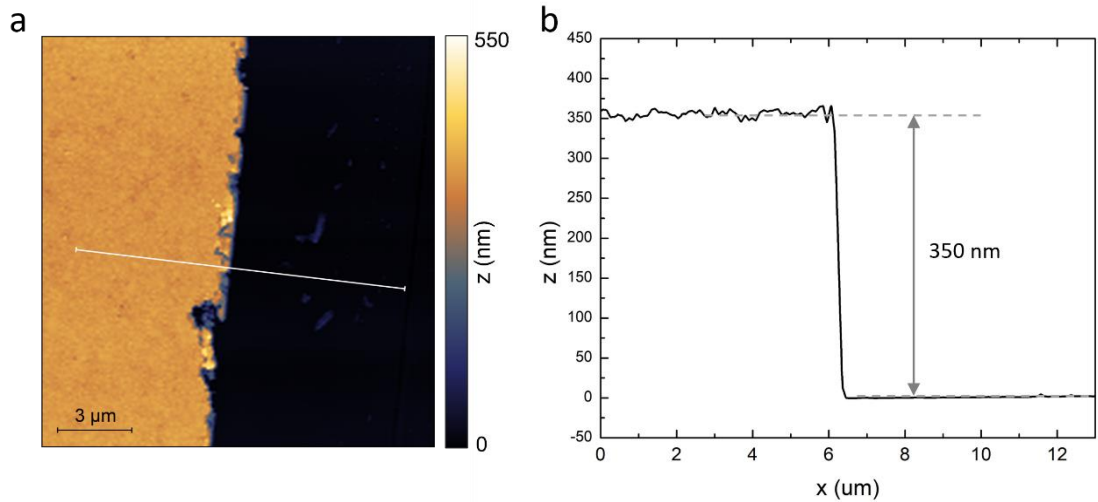

**Figure S13. Atomic force microscopy (AFM) characterization of MAPbI$_3$ film.** a) AFM image of the perovskite film on a scratched edge. b) AFM profile of the path indicated by the white line in a), showing a film thickness of 350 nm.